\documentclass[pra, reprint, twocolumn, superscriptaddress, amsmath, amssymb, aps, ,floatfix]{revtex4-2}

\usepackage[english]{babel}
\usepackage{graphicx}
\graphicspath{{./Images/},{./ImagesAppendix/},{./images/},{./imagesAppendix/}}

\usepackage{amsmath, amssymb, amsfonts}
\usepackage{physics}
\usepackage{bm}
\usepackage{braket}
\usepackage{array}
\usepackage{multirow}
\usepackage{float}
\usepackage{commath}
\usepackage{verbatim}
\usepackage{color}
\usepackage{enumitem}
\usepackage{array}   
\usepackage{dsfont}
\usepackage{mathtools}
\usepackage{soul}
\usepackage[colorlinks=true, citecolor=darkBlue, linkcolor=darkBlue, urlcolor=blue]{hyperref}
\usepackage{amssymb}
\usepackage{tabularx}
\usepackage{booktabs}
\definecolor{darkBlue}{rgb}
{0.08, 0.13, 0.4}

\usepackage{cancel}

\usepackage{tikz}
\usetikzlibrary{positioning, arrows.meta}
\usepackage{pgfplots}

\definecolor{THc}{rgb}{0.9,0.3,0.2}

\newcommand{\canc}[1]{}

\begin{document}

\title{Fock-space delocalization and its control in scarred-vs-ergodic dynamics}
\author{Francesco Perciavalle}
\affiliation{Dipartimento di Fisica, Universit\`a della Calabria, 87036 Arcavacata di Rende (CS), Italy}
\affiliation{INFN, Sezione LNF, gruppo collegato di Cosenza}

\date{\today}

\begin{abstract}
Quantum many-body scarred dynamics in Rydberg-atom arrays is characterized by revivals, in which the system repeatedly returns to its initial state despite its many-body complexity. But how extensively does the wave function explore the space of allowed configurations between revivals? We address this question by quantifying the fraction of configurations occupied during PXP dynamics initialized from the paradigmatic polarized and Néel states, which respectively exhibit thermalizing and scarred dynamics. Despite their striking revivals, dynamics from the Néel state spreads over exponentially many configurations, yet occupies a vanishing fraction of the full space in the thermodynamic limit. Dynamics from the polarized state, by contrast, occupies a finite fraction of the configuration space, close to anticoncentration and random-state statistics, although chiral symmetry limits this fraction to one third. A weak quasiperiodic field enables the polarized state to occupy half of the configuration space by lifting the constraint imposed by chiral symmetry. The Néel state, however, requires a stronger field to reach a comparable anticoncentrated regime. As the field increases further, the occupied fraction decreases for both initial states, revealing a nonmonotonic dependence of Fock-space delocalization on field strength without necessarily implying localization.
\end{abstract}

\maketitle

\section{Introduction}
Arrays of neutral atoms excited to Rydberg states are among the most
versatile platforms for the quantum simulation of many-body
dynamics~\cite{bernien2017probing, browaeys2020many, wu2021concise, adams2019rydberg, bai2026many}.
In a one-dimensional chain, the Rydberg blockade~\cite{gatan2009observation,urban2009observation,kitson2024rydberg} forbids two adjacent atoms
from being simultaneously excited, and the dynamics is captured by the PXP
model~\cite{turner2018quantum, serbyn2021quantum, lesanovsky2012interacting, park2025nonintegrability}, a non-integrable model
whose accessible Hilbert space has dimension given by the Fibonacci number
$F_{L+2}$ rather than $2^L$. Although this model is expected to thermalize
according to the eigenstate thermalization
hypothesis (ETH)~\cite{deutsch2018eigenstate, srednicki1994chaos,
dalessio2016from, rigol2008themralization}, experiments revealed a striking
dependence on the initial state~\cite{bernien2017probing}: the N\'eel state
$\ket{\mathbb Z_2}=\ket{1010\ldots}$ displays long-lived revivals, whereas
the polarized state $\ket{\mathbf 0}=\ket{00\ldots0}$ relaxes quickly. The
revivals originate from quantum many-body scars (QMBS), a small set of
non-thermal eigenstates, approximately equally spaced in energy, with
anomalously large overlap with $\ket{\mathbb Z_2}$~\cite{perciavalle2026floquet,turner2018quantum,
serbyn2021quantum, moudgalya2022quantum, choi2019emergent}. Scarred
dynamics has since been found in several models and several
platforms~\cite{liang2025observation, zhang2022many, zhao2020quantum,
kaneko2024quantum, iadecola2020quantum, ebner2026global, osborne2026quantum, aramthottil2022scar, gotta2023asymtotic, logaric2024quantum} and is a paradigmatic example of weak ergodicity
breaking.

\begin{figure}[!t]
    \centering
    \includegraphics[width=\linewidth]{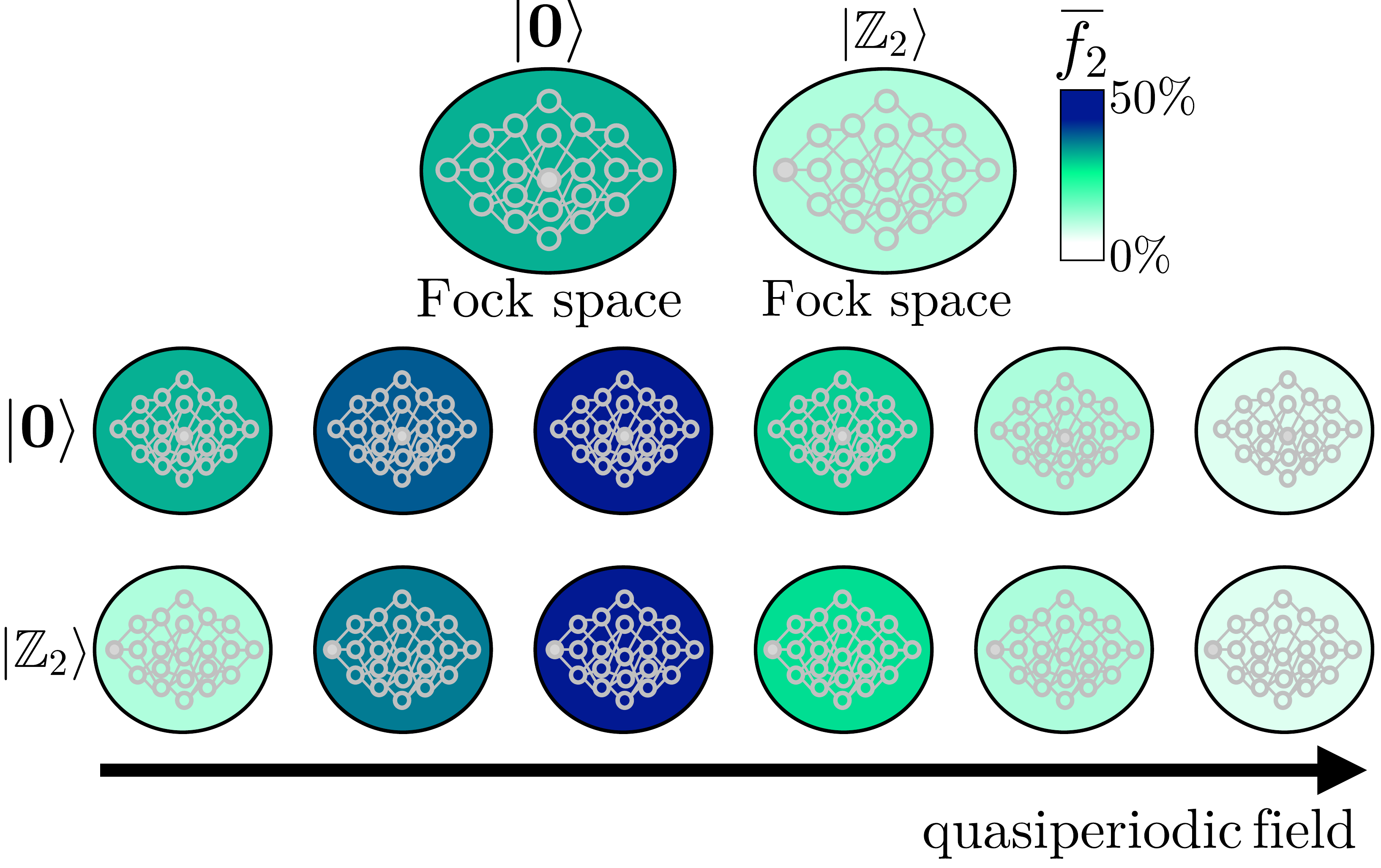}
\caption{How far does a quantum state spread in Fock space? Each ellipse
represents the Fock space of a Rydberg-atom chain: dots are the allowed
configurations ($L=6$), and lines connect configurations linked directly
by the dynamics. The gray dot marks the initial configuration, and the
color shows $\overline{f}_2$, the time-averaged fraction of the Fock space
effectively occupied by the wave function, the choice of colors is schematic.
Top: without field, the polarized state $\ket{\mathbf 0}$ spreads over a
large part of the Fock space, while the scarred N\'eel state
$\ket{\mathbb Z_2}$ occupies a smaller, but still sizable, fraction.
Bottom: a quasiperiodic field first enhances the spreading of both states,
up to about half of the Fock space, and then suppresses it.}
    \label{fig:sketch}
\end{figure}
Scarred and thermalizing dynamics are commonly distinguished by their ability to retain memory of the initial state, as diagnosed by the return probability or the relaxation of local observables~\cite{turner2018quantum, serbyn2021quantum}. For an initial product state, the wave function is initially localized on a single atomic configuration, and the return probability measures the weight that remains on this configuration during time evolution. As this memory is lost, the weight spreads over other configurations. However, while conventional diagnostics quantify the extent of memory loss, they do not reveal how the wave-function weight is redistributed.
To track this redistribution, we turn to the \emph{blockaded Fock space} (BFS), the set of atomic configurations compatible with the blockade constraint. In this basis, the PXP Hamiltonian defines a hopping problem on a graph, where each configuration is connected to those accessible through a single-atom flip. The dynamics can thus be viewed as the spreading of the initial state's weight across this graph (Fig.~\ref{fig:sketch}). Characterizing this spreading in full is a formidable task. Here, we take a first step by asking a simpler question: \textit{what fraction of the BFS does the wave function explore?} The pronounced revivals characteristic of scarred dynamics suggest that the wave function remains concentrated on a small subset of configurations, whereas thermalizing dynamics might be expected to explore a much larger portion of the BFS. Testing this expectation amounts to probing the extent of delocalization in Fock space.

The question of how many configurations a quantum state occupies has a long history in the study of many-body localization~\cite{altshuler1997quasiparticle, basko2006metal, luitz2015many, mace2019multifractal}. Participation entropies and fractal dimensions provide a quantitative way to distinguish localized, extended, and multifractal states~\cite{luitz2014universal, janssen1994multifractal, evers2008anderson}. A closely related question arises in quantum computation through the concept of \emph{anticoncentration}: a state is anticoncentrated when measurement outcomes in the computational basis are broadly distributed across configurations, rather than concentrated on a few~\cite{dalzell2022random, hangleiter2018anticoncentration}. Anticoncentration is a key ingredient in sampling-based quantum advantage~\cite{hangleiter2023computational, heinrich2026anticoncentration, hangleiter2018anticoncentration} and has been studied extensively in random circuits and ergodic many-body dynamics~\cite{dalzell2022random, claeys2025fock, turkeshi2024hilbert, tirrito2025anticoncentration, aditya2026coherence}. Delocalization, however, can occur to different degrees. A state is localized in Fock space if the number of significantly occupied configurations remains bounded as the system size increases. If this number grows with system size but remains subextensive relative to the Hilbert-space dimension, the state is delocalized but occupies a vanishing fraction of the available configurations. By contrast, if the number of occupied configurations scales with the Hilbert-space dimension, the state occupies a finite fraction of the space and exhibits anticoncentration~\cite{tirrito2025anticoncentration, deshpande2022tight}. The precise fraction depends on the statistics of the wave-function amplitudes: random states occupy, on average, one-third of the configurations for real amplitudes and one-half for complex amplitudes~\cite{Lakshminarayan2026}. How scarred dynamics fits into this picture remains an open question. Addressing it is also experimentally relevant: measurements of individual atomic occupations in Rydberg quantum simulators yield precisely the Fock-space configurations and probabilities needed to characterize this spreading~\cite{bernien2017probing}.

Delocalization can also be probed through the response to perturbations. In tight-binding models, quasiperiodic potentials provide a paradigmatic mechanism for localization: in the Aubry--André model, all eigenstates become localized above a critical potential strength~\cite{roati2008anderson, modugno2009exponential, hetenyi2025numerical, longhi2021phase, an2021interactions}. Since the PXP model can itself be viewed as a hopping problem on the BFS graph, it is natural to ask how its Fock-space dynamics responds to a quasiperiodic longitudinal field. Such a field detunes the graph's configurations and, at the same time, breaks the model's chiral symmetry, allowing us to probe the interplay between Fock-space delocalization and quantum scars.

Here we show that scarred dynamics is delocalized in Fock space but remains far from ergodic. Both the scar eigenstates and the dynamics initiated from $\ket{\mathbb Z_2}$ explore an exponentially growing number of configurations, yet occupy an exponentially vanishing fraction of the BFS. By contrast, dynamics initiated from $\ket{\mathbf 0}$ spreads across the BFS with random-state statistics approaching anticoncentration. However, because chiral symmetry constrains its amplitudes to be real or purely imaginary, it occupies only about one-third of the BFS, rather than the one-half expected for random states with complex amplitudes.

The quasiperiodic field reveals how these two effects compete. A weak field breaks chiral symmetry, bringing the fraction of the BFS occupied by the dynamics from $\ket{\mathbf 0}$ to one-half. For $\ket{\mathbb Z_2}$, in contrast, this fraction reaches one-half and anticoncentration only when the field becomes comparable to the PXP energy scale, where the scars are destroyed. In this regime, $\ket{\mathbf 0}$ and $\ket{\mathbb Z_2}$ initial states become indistinguishable in their Fock-space spreading. At still stronger fields, spreading is suppressed through a crossover rather than a sharp transition: the dynamics continues to explore exponentially many configurations, but their fraction of the BFS vanishes, recovering the scaling found for scarred dynamics in the absence of a field.

\section{Model and Methods}
\label{sec:model}
We consider a chain of $L$ Rydberg atoms in the nearest-neighbor blockade
regime, described by the PXP model~\cite{bernien2017probing,
turner2018quantum, serbyn2021quantum},
\begin{equation}
    \hat{H}_{\rm PXP}
    =
    \Omega\sum_{j=1}^{L}
    \hat{P}_{j-1}\hat{X}_j\hat{P}_{j+1},
    \label{eq:HPXP}
\end{equation}
with open boundary conditions (OBC), $\hat P_0=\hat P_{L+1}\equiv\hat{\mathds 1}$.
Here, $\hat{X}_j=\ket{1}_j\bra{0}+\ket{0}_j\bra{1}$ and
$\hat{Z}_j=\ket{1}_j\bra{1}-\ket{0}_j\bra{0}$ are Pauli matrices,
$\hat{P}_j=\frac{1}{2}(\hat{\mathds{1}}_j-\hat{Z}_j)$ is the projector onto
the ground state $\ket{0}_j$, and $\hat{N}_j=\ket{1}_j\bra{1}$ is the local
excitation number. An atom can thus be excited or de-excited only if both
its neighbors are in the ground state. We set $\hbar=1$ and $\Omega=1$, so that times are
expressed in units of $\Omega^{-1}$.

The dynamics conserves the blockade constraint: an initial state with no
two adjacent excitations keeps this property at all times. The evolved
state can therefore be expanded on the blockaded Fock space (BFS),
\begin{equation}
    \ket{\psi(t)}
    =
    e^{-i\hat{H}_{\rm PXP}t}\ket{\psi(0)}
    =
    \sum_{\mathbf{n}\in\mathcal{B}_L}
    C_{\mathbf{n}}(t)\ket{\mathbf{n}},
    \label{eq:expansion}
\end{equation}
where $\mathcal{B}_L=\{\ket{\mathbf{n}}=\ket{n_1,\ldots,n_L}:
n_jn_{j+1}=0\}$, $C_{\mathbf{n}}(t)=\braket{\mathbf{n}|\psi(t)}$. Its dimension is the Fibonacci number
$d_{\mathcal{B}}=F_{L+2}\approx\varphi^{L+2}/\sqrt{5}$, with
$\varphi=(1+\sqrt{5})/2$ being the golden ratio. The BFS is thus exponentially
large in $L$, but grows more slowly than the full Hilbert space:
$\ln d_{\mathcal B}\simeq L\ln\varphi+c_0$, with $\ln\varphi\simeq0.48$
and $c_0=2\ln\varphi-\tfrac12\ln5\simeq0.16$, instead of $L\ln2$.
In this basis, $\hat H_{\rm PXP}$ has no diagonal elements, and its
off-diagonal elements connect BFS states that differ by a single allowed
spin flip. The PXP model can thus be viewed as a hopping problem for a
single fictitious particle moving on a graph whose vertices are the BFS
states~\cite{turner2018quantum},
\begin{equation}
    \hat H_{\rm PXP}
    =
    \Omega\sum_{\mathbf n,\mathbf m\in\mathcal B_L}
    A_{\mathbf n\mathbf m}\,\ket{\mathbf n}\bra{\mathbf m},
    \label{eq:hopping}
\end{equation}
where the adjacency matrix $A_{\mathbf n\mathbf m}=1$ if $\mathbf m$ is
obtained from $\mathbf n$ by flipping a single atom $j$ with
$n_{j-1}=n_{j+1}=0$, and $A_{\mathbf n\mathbf m}=0$ otherwise. All
hoppings are equal and there are no on-site energies: the only structure
comes from the graph itself, whose connectivity varies across the BFS.
In Fig.~\ref{fig:sketch} we report in each ellipse examples of graphs 
for the PXP model with OBC and $L=6$.
The time evolution of Eq.~\eqref{eq:expansion} is then the spreading of
this particle over the graph, and the participation entropies introduced
below quantify how far it has spread.

\paragraph*{Participation entropies.}
We characterize how the weight $p_{\mathbf n}(t)=|C_{\mathbf n}(t)|^2$ is
spread over the BFS through the participation
entropies~\cite{luitz2014universal, claeys2025fock,
tirrito2025anticoncentration, aditya2026coherence}
\begin{equation}
    S_q(t)=\frac{1}{1-q}\ln I_q(t),
    \qquad
    I_q(t)=\sum_{\mathbf n\in\mathcal B_L}p_{\mathbf n}(t)^{q},
    \label{eq:Sq}
\end{equation}
i.e., the R\'enyi entropies of the distribution $p_{\mathbf n}$, with
$I_q$ the generalized inverse participation ratio. The index $q$ selects
which part of the distribution is probed. For $q>1$ the largest
probabilities dominate, while for $q<1$ small but nonzero probabilities
weigh more. The limit $q\to1$ gives the Shannon entropy
$S_1=-\sum_{\mathbf n}p_{\mathbf n}\ln p_{\mathbf n}$. For $q=2$, the
participation ratio $\mathrm{PR}_2=I_2^{-1}=e^{S_2}$ gives the effective
number of BFS states occupied by the wave
function~\cite{evers2008anderson}. It ranges from $\mathrm{PR}_2=1$ for a
single Fock state to $\mathrm{PR}_2=d_{\mathcal B}$ for the uniform
superposition over the whole BFS, for which $S_q=\ln d_{\mathcal B}$ for
every $q$.

\paragraph*{Participation fraction and population fluctuations.}
The fraction of the BFS effectively occupied is the participation
fraction
\begin{equation}
    f_2=\frac{\mathrm{PR}_2}{d_{\mathcal B}}=e^{S_2-\ln d_{\mathcal B}}.
    \label{eq:fPR}
\end{equation}
It is directly related to the fluctuations of the populations across the
BFS. Since $\sum_{\mathbf n}p_{\mathbf n}=1$, the populations have mean
$\overline p=1/d_{\mathcal B}$, and we define their variance as
\begin{equation}
    \sigma_p^2=\frac{1}{d_{\mathcal B}}
    \sum_{\mathbf n\in\mathcal B_L}
    \left(p_{\mathbf n}-\frac{1}{d_{\mathcal B}}\right)^2 .
    \label{eq:population_variance}
\end{equation}
Expanding the square gives
$I_2=(1+\sigma_p^2d_{\mathcal B}^2)/d_{\mathcal B}$, and therefore
\begin{equation}
    \mathrm{PR}_2=\frac{d_{\mathcal B}}{1+\sigma_p^2d_{\mathcal B}^2},
    \qquad
    f_2=\frac{1}{1+\sigma_p^2d_{\mathcal B}^2}.
    \label{eq:fPR_variance}
\end{equation}
Small relative fluctuations of the populations, $\sigma_p d_{\mathcal B}$,
thus correspond to a weight spread over many BFS states, and large ones to
a weight concentrated on a few of them. In particular, $f_2$ remains finite
only if $\sigma_p d_{\mathcal B}$ stays bounded as the system size
increases. This is the condition of \emph{anticoncentration}: a state is anticoncentrated when
$I_2\le 1/(\alpha d_{\mathcal B})$ ($0<\alpha\leq 1$) for some constant
$\alpha$~\cite{tirrito2025anticoncentration,dalzell2022random, hangleiter2018anticoncentration}, i.e.,
when $f_2\ge \alpha$ remains finite as $L\to\infty$.

\paragraph*{Fractal dimension.}
The extent of a state in Fock space is set by how $S_2$ scales with
$\ln d_{\mathcal B}$, through the fractal
dimension~\cite{luitz2014universal, hamanaka2025multifractality}
\begin{equation}
    \mathcal D_2=\lim_{d_{\mathcal B}\to\infty}
    \frac{S_2}{\ln d_{\mathcal B}},
    \label{eq:fractal}
\end{equation}
which measures how the number of effectively occupied basis states grows
with the size of the Hilbert space. Since $d_{\mathcal B}\sim\varphi^L$
with $\varphi=(1+\sqrt5)/2$, one has $\ln d_{\mathcal B}\simeq L\ln\varphi$,
and a linear growth $S_2\simeq aL$ gives
\begin{equation}
    \mathcal D_2\simeq\frac{a}{\ln\varphi},
    \qquad
    f_2\sim e^{-(1-\mathcal D_2)L\ln\varphi}=e^{-(\ln\varphi-a)L}.
    \label{eq:f_scaling}
\end{equation}
For a localized state, $\mathcal D_2=0$: $\mathrm{PR}_2$ remains bounded as
$L\to\infty$. For an extended state, $\mathcal D_2=1$: $\mathrm{PR}_2$
grows proportionally to $d_{\mathcal B}$ at leading exponential order. Intermediate values, $0<\mathcal D_2<1$, indicate
that $\mathrm{PR}_2$ grows exponentially with $L$ but more slowly than
$d_{\mathcal B}$. Such states occupy exponentially many basis states, but
an exponentially vanishing fraction of the BFS. The scaling of $f_2$
therefore provides a direct diagnostic of Fock-space delocalization, while
comparison with the appropriate random-state reference is needed to assess
ergodicity. In particular, an anticoncentrated state is characterized by a finite occupied
fraction, $f_2=\alpha$, with $0<\alpha\leq 1$ independent of $L$
(equivalently, $\mathrm{PR}_2=\alpha d_{\mathcal B}$). Consequently,
$S_2=\ln\mathrm{PR}_2=\ln d_{\mathcal B}+\ln\alpha$, so that
$S_2/\ln d_{\mathcal B}\to 1$ as $L\to\infty$, yielding $\mathcal D_2=1$.

\begin{figure*}[!t]
    \centering
    \includegraphics[width=\linewidth]{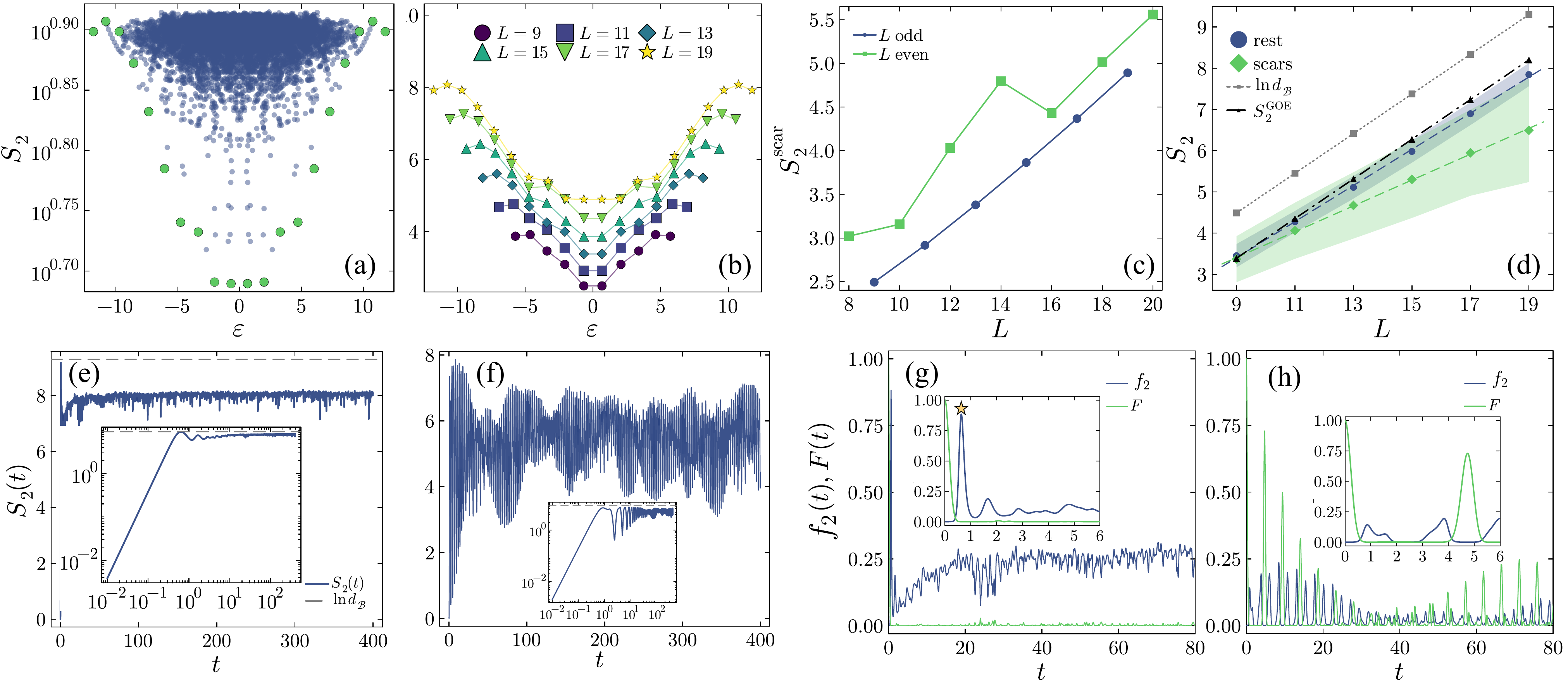}
\caption{Participation entropy $S_2$ in the spectrum and Fock-space dynamics of the PXP model with OBC.
(a) $S_2$ of the eigenstates $\ket{\varepsilon}$ versus $\varepsilon$ for $L=19$: the $L+1$ scars (green), identified from the upper arc of the overlap with $\ket{\mathbb Z_2}$.
(b) $S_2$ of the scars for $L=9,\dots,19$.
(c) $S_2$ of the central scar (closest to $\varepsilon=0$, $\varepsilon\neq0$) versus $L$, for odd and even sizes.
(d) Mean $S_2$ of scars (diamonds) and of all other eigenstates (circles) versus $L$, with standard deviation bands and linear fits $S_2^{\rm scar}\simeq0.31\,L+0.58$ and $S_2^{\rm rest}\simeq0.44\,L-0.54$. Triangles show the random-state value $S_2^{\rm GOE}=\ln d_{\mathcal B}-\ln3$ (Appendix~\ref{app:ergodic}) and squares the maximum $\ln d_{\mathcal B}$.
(e),(f) $S_2(t)$ for $L=19$ starting from $\ket{\mathbf 0}$ (e) and $\ket{\mathbb Z_2}$ (f); the dashed line marks $\ln d_{\mathcal B}$. The insets (log-log) show the quadratic growth at short times.
(g),(h) $f_2(t)$ (blue) and fidelity $F(t)=|\langle\psi(0)|\psi(t)\rangle|^2$ (green) for the same initial states. The insets zoom in on short times; the star marks the transient maximum $f_2\simeq0.88$. The ergodic reference is $f_2^{\rm GOE}=1/3$.}
\label{fig:spectrum_dynamics}
\end{figure*}
\paragraph*{Ergodic reference.}
For an ergodic state, $f_2$ depends on the statistics of the amplitudes $C_{\mathbf n}$~\cite{Lakshminarayan2026, torres2016realistic, schiulaz2018from} 
and hence on the symmetries of the system. The PXP Hamiltonian is real and
anticommutes with the chiral operator $\hat{\mathcal C}=\prod_j\hat Z_j$,
which constrains the amplitudes $C_{\mathbf n}(t)$ evolved from a Fock
state to be either real or purely imaginary (Appendix~\ref{app:ergodic}).
The appropriate benchmark is therefore that of random real Gaussian
amplitudes, as in the Gaussian Orthogonal Ensemble (GOE), which in the
thermodynamic limit gives the anticoncentrated values~\cite{Lakshminarayan2026}
\begin{equation}
    f_2^{\rm GOE}\simeq\frac13,
    \qquad
    S_2^{\rm GOE}\simeq\ln d_{\mathcal B}-\ln3.
    \label{eq:GOE_ergodic}
\end{equation}
Random states with complex Gaussian amplitudes, as in the Gaussian Unitary
Ensemble (GUE) or for Haar-random states, instead
give~\cite{Lakshminarayan2026}
\begin{equation}
    f_2^{\rm GUE}\simeq\frac12,
    \qquad
    S_2^{\rm GUE}\simeq\ln d_{\mathcal B}-\ln2.
    \label{eq:GUE_ergodic}
\end{equation}
The generalization to arbitrary $S_q$ is given in
Appendix~\ref{app:ergodic_q}. Here GOE and GUE refer to the statistics of
the wave-function amplitudes, not to the symmetry class of the Hamiltonian.
Since $\hat H_{\rm PXP}$ is real, its eigenstates can be chosen real and are
benchmarked against GOE. Reality alone, however, does not make the
time-evolved amplitudes real: it is the chiral symmetry that constrains them
to be real or purely imaginary. In Sec.~\ref{sec:qp}, we add a longitudinal
field that keeps the Hamiltonian real but breaks the chiral symmetry. The
eigenstates then still refer to GOE statistics, whereas the time-evolved
amplitudes acquire independent dynamical phases, and the relevant reference
for the dynamics changes from $f_2^{\rm GOE}\simeq1/3$ to
$f_2^{\rm GUE}\simeq1/2$~\cite{Lakshminarayan2026, claeys2025fock, tirrito2025anticoncentration}.

\section{Fock-space delocalization}
\label{sec:spectral_dynamics}
We now apply the tools introduced above to characterize how quantum
states spread over the BFS. We use the participation entropy $S_2$ of
Eq.~(\ref{eq:Sq}) and the participation fraction $f_2$ of
Eq.~(\ref{eq:fPR}), and classify the states through the fractal dimension
$\mathcal D_2$ of Eq.~(\ref{eq:fractal}), using the GOE value of
Eq.~(\ref{eq:GOE_ergodic}) as the ergodic reference. We first consider the
eigenstates of $\hat H_{\rm PXP}$, focusing on the contrast between the
scars and the rest of the spectrum. We then turn to the dynamics of
initial Fock states and ask whether it reflects the spectral picture.

\subsection{Spectral Fock-space delocalization}
\label{sec:spectral}
For the eigenstates
$\hat{H}_{\rm PXP}\ket{\varepsilon}=\varepsilon\ket{\varepsilon}$, the
populations are $p_{\mathbf n}=|\langle\mathbf n|\varepsilon\rangle|^2$.
As shown in Fig.~\ref{fig:spectrum_dynamics}(a), the behavior of the scars
depends on their position in the spectrum. Scars at the edges of the
spectrum have values of $S_2$ comparable to those of nearby eigenstates.
Scars near the center, which have the largest overlap with
$\ket{\mathbb Z_2}$, have a markedly smaller $S_2$.

A small $S_2$ at fixed $L$, however, does not imply localization, which
according to Eq.~(\ref{eq:fractal}) is set by the scaling of $S_2$ with
system size. For every scar, $S_2$ grows approximately linearly with $L$
[Fig.~\ref{fig:spectrum_dynamics}(b)], so that $\mathrm{PR}_2=e^{S_2}$
grows exponentially and $\mathcal D_2>0$: the scars are not localized.
This holds even for the most concentrated scar, the central one
[Fig.~\ref{fig:spectrum_dynamics}(c)]. Its growth is almost perfectly
linear for odd $L$ and more irregular for even $L$, but shows no sign of
saturation.

The delocalized nature of the scars is encoded in how fast
$S_2$ grows compared with $\ln d_{\mathcal B}\simeq L\ln\varphi$.
Fig.~\ref{fig:spectrum_dynamics}(d) shows that the average $S_2$ of both
the scars and the other eigenstates grows linearly, $S_2\simeq aL+b$, so
that Eq.~(\ref{eq:f_scaling}) directly gives $\mathcal D_2\simeq a/\ln\varphi$
and $f_2\sim e^{-(\ln\varphi-a)L}$. The occupied fraction therefore stays
finite only if $a=\ln\varphi$, i.e., $\mathcal D_2=1$.

The non-scar eigenstates closely follow the GOE reference
$S_2^{\rm GOE}\simeq\ln d_{\mathcal B}-\ln3$ of Eq.~(\ref{eq:GOE_ergodic}),
which provides the appropriate ergodic benchmark for the PXP Hamiltonian.
Their slope, $a=0.44$, yields $\ln\varphi-a\simeq0.04$ and
$\mathcal D_2\simeq0.91$. Thus, although their fractal dimension is
slightly below unity, these states are close to being fully delocalized
in the BFS, with a participation fraction that decays only marginally
with system size.

The scars, by contrast, have a smaller slope, $a=0.31$, corresponding
to $\mathcal D_2\simeq0.64$. Their participation fraction therefore
decays as $e^{-0.17L}$. While the scars also occupy exponentially many
BFS states, they occupy an exponentially vanishing fraction of the
BFS, reflecting a substantially stronger departure from full
delocalization than that of the non-scar eigenstates. This
average hides a strong energy dependence, reflected in the large standard
deviation over the scars: the edge scars are close to the rest of the
spectrum, while the central ones are the most concentrated.

One may expect the dynamics to inherit these features. If $\ket{\psi(0)}$
lies mostly on the scars $\ket{\mathcal S_\alpha}$, as the N\'eel state
does, then
\begin{equation}
    \ket{\psi(t)}\simeq\sum_{\alpha=1}^{L+1}\eta_\alpha\,
    e^{-i\varepsilon_\alpha t}\ket{\mathcal S_\alpha},
    \qquad \eta_\alpha=\langle\mathcal S_\alpha|\psi(0)\rangle,
    \label{eq:scar_superposition}
\end{equation}
and its spreading over the BFS should reflect the fractal dimension of the
scars, occupying an exponentially vanishing
fraction of the BFS. This is not
guaranteed, however, because participation entropies are nonlinear in the
state. Two states each peaked on the same configuration can combine, once
the peaks cancel, into a state spread over the whole BFS through their
exponentially small tails, which the participation ratio does not detect.
The delocalization of the dynamics must therefore be addressed directly.

\subsection{Dynamical Fock-space delocalization}
\label{sec:dynamics}
We follow the spreading over the BFS of an initial Fock state
$\ket{\mathbf n_0}$, for which $S_q(0)=0$ and $f_2(0)=1/d_{\mathcal B}$,
comparing the polarized state $\ket{\mathbf 0}$ (ergodic) with the N\'eel
state $\ket{\mathbb Z_2}$ (scarred).
For $t\ll\frac{2}{\sqrt q}\frac{\tau_{\rm MT}}{\pi}$, the participation
entropies grow quadratically,
\begin{equation}
    S_q(t)\simeq\frac{q}{q-1}(\Delta H_{\rm PXP})^2t^2
    \qquad(q>1),
    \label{eq:short_time}
\end{equation}
where $\tau_{\rm MT}=\pi/(2\Delta H_{\rm PXP})$ is the Mandelstam--Tamm
quantum speed limit (Appendix~\ref{app:short-time})~\cite{deffner2017quantum,
perciavalle2024quantum, mandelstam1991uncertainty, giovannetti2003quantum}.
Since all hoppings in Eq.~(\ref{eq:hopping}) are equal,
$(\Delta H_{\rm PXP})^2=\sum_{\mathbf n\neq\mathbf n_0}
    |\langle\mathbf n|\hat H_{\rm PXP}|\mathbf n_0\rangle|^2$ is directly related to the number of links of the initial state:
$L$ for $\ket{\mathbf 0}$ and about $L/2$ for $\ket{\mathbb Z_2}$. The BFS
connectivity therefore sets the initial growth [insets of
Fig.~\ref{fig:spectrum_dynamics}(e),(f)]. The differences between the two
states emerge beyond this regime.

\paragraph*{Polarized state.}
After the quadratic regime, $S_2(t)$ quickly reaches a plateau below
$\ln d_{\mathcal B}$ and then drifts only slowly
[Fig.~\ref{fig:spectrum_dynamics}(e)]. Correspondingly, $f_2$ remains below
the ergodic value $f_2^{\rm GOE}\simeq1/3$ but slowly approaches it
[Fig.~\ref{fig:spectrum_dynamics}(g)], consistent with the delocalized
character of the non-scar eigenstates. At short times, while $F(t)$ drops,
$f_2$ shows a transient maximum at which $88\%$ of the BFS is effectively
occupied, well above the ergodic value, in line with
Ref.~\cite{claeys2025fock}. By Eq.~(\ref{eq:fPR_variance}), this
corresponds to nearly uniform populations, with low relative fluctuations
$\sigma_p d_{\mathcal B}\simeq0.36$, small compared with $\sqrt2$ for GOE random
states. This nearly uniform occupation is not sustained: population
fluctuations then grow and $f_2$ decreases.

\paragraph*{N\'eel state.}
$S_2(t)$ oscillates strongly [Fig.~\ref{fig:spectrum_dynamics}(f)]. It
returns close to zero at the revivals of $F(t)$, reflecting the coherent
oscillations between $\ket{\mathbb Z_2}$ and $\ket{\mathbb Z_2'}$. As the
revivals decay, its minima rise, signaling progressive spreading over the
BFS. The occupied fraction $f_2$ is anticorrelated with $F(t)$
[Fig.~\ref{fig:spectrum_dynamics}(h)]. Between consecutive revivals it
shows two peaks separated by a minimum at half period, when the state
passes near $\ket{\mathbb Z_2'}$. This agrees with the SU(2)
picture~\cite{choi2019emergent}: midway between the two N\'eel states, the
effective large spin lies on the equator and is spread over many
configurations. The peaks reach $f_2\simeq25\%$, a sizable fraction of
$f_2^{\rm GOE}\simeq1/3$. Even in the scarred regime, therefore, the wave
function periodically occupies a substantial portion of the BFS.

\section{Scaling analysis}
\begin{figure}[!t]
    \centering
    \includegraphics[width=\linewidth]{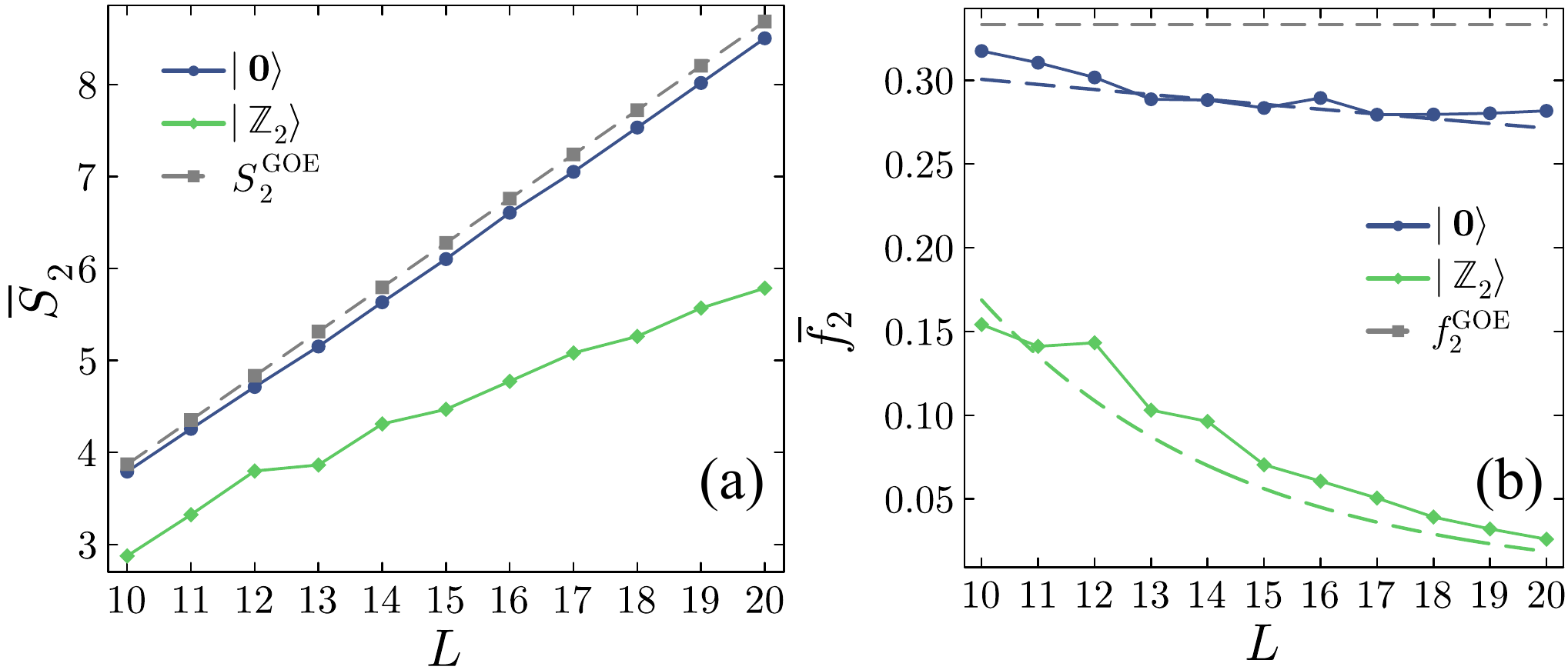}
\caption{Long-time scaling of the participation entropy and of the
participation fraction with system size $L$. All quantities are time
averages up to $t=800$.
(a) $\overline S_2$ versus $L$ for the polarized state $\ket{\mathbf 0}$
and the N\'eel state $\ket{\mathbb Z_2}$, compared with the ergodic
reference $S_2^{\rm GOE}=\ln d_{\mathcal B}-\ln3$
[Eq.~(\ref{eq:GOE_ergodic})]; dashed lines are linear fits.
(b) $\overline f_2$ versus $L$ (solid lines). The colored dashed lines show
the lower bound of Eq.~(\ref{eq:fPR_scaling}), evaluated with the linear
fits of (a); the gray dashed line marks $f_2^{\rm GOE}=1/3$.}
    \label{fig:scaling}
\end{figure}
To determine which kind of Fock-space delocalization the two initial
states realize, we consider the long-time averages
\begin{equation}
\overline{X}=\frac{1}{t_f}\int_0^{t_f} dt'\,X(t'),  \,\,\,\,
X=S_2,f_2
\end{equation}
evaluated at
$t_f=800$. For $\ket{\mathbf 0}$, this window extends well beyond the
relaxation discussed above, so that $\overline X$ characterizes the
long-time state. For $\ket{\mathbb Z_2}$, the time average must be
interpreted with more care. As shown in
Fig.~\ref{fig:spectrum_dynamics}(f),(h), the dynamics remains strongly
oscillating: near the revivals the wave function is concentrated on a few
configurations close to $\ket{\mathbb Z_2}$ or $\ket{\mathbb Z_2'}$, while
at intermediate times it spreads over a sizable portion of the BFS. In this
case, $\overline X$ does not describe a stationary state, but the structure
of the wave function averaged over many revival periods, and all
statements below about the N\'eel dynamics should be understood in this
averaged sense.

Figure~\ref{fig:scaling}(a) shows $\overline S_2$ as a function of $L$ for
both initial states, together with the ergodic reference
$S_2^{\rm GOE}$ of Eq.~(\ref{eq:GOE_ergodic}). For $\ket{\mathbf 0}$,
$\overline S_2$ grows linearly with $L$ and runs parallel to the
reference, just below it: a linear fit $\overline S_2\simeq aL+b$ gives
$a=0.471$, compatible with the maximal slope $\ln\varphi\simeq0.481$,
i.e., $\overline{\mathcal D}_2\simeq0.98$. For $\ket{\mathbb Z_2}$, $\overline S_2$
also increases with $L$, but less regularly, with even-odd effects
reminiscent of Fig.~\ref{fig:spectrum_dynamics}(c), and with a markedly
smaller slope, $a=0.261$ for $L\geq15$, i.e., $\ln\varphi-a\simeq0.22$ and
$\overline{\mathcal D}_2\simeq0.54$. In both cases $\mathrm{PR}_2$ grows
exponentially with $L$, so that neither dynamics is localized, even in the
scarred regime. The N\'eel dynamics thus inherits the behavior of the
scars found in Sec.~\ref{sec:spectral}, with an even smaller slope $a$.

\begin{figure*}[!t]
    \centering
    \includegraphics[width=\linewidth]{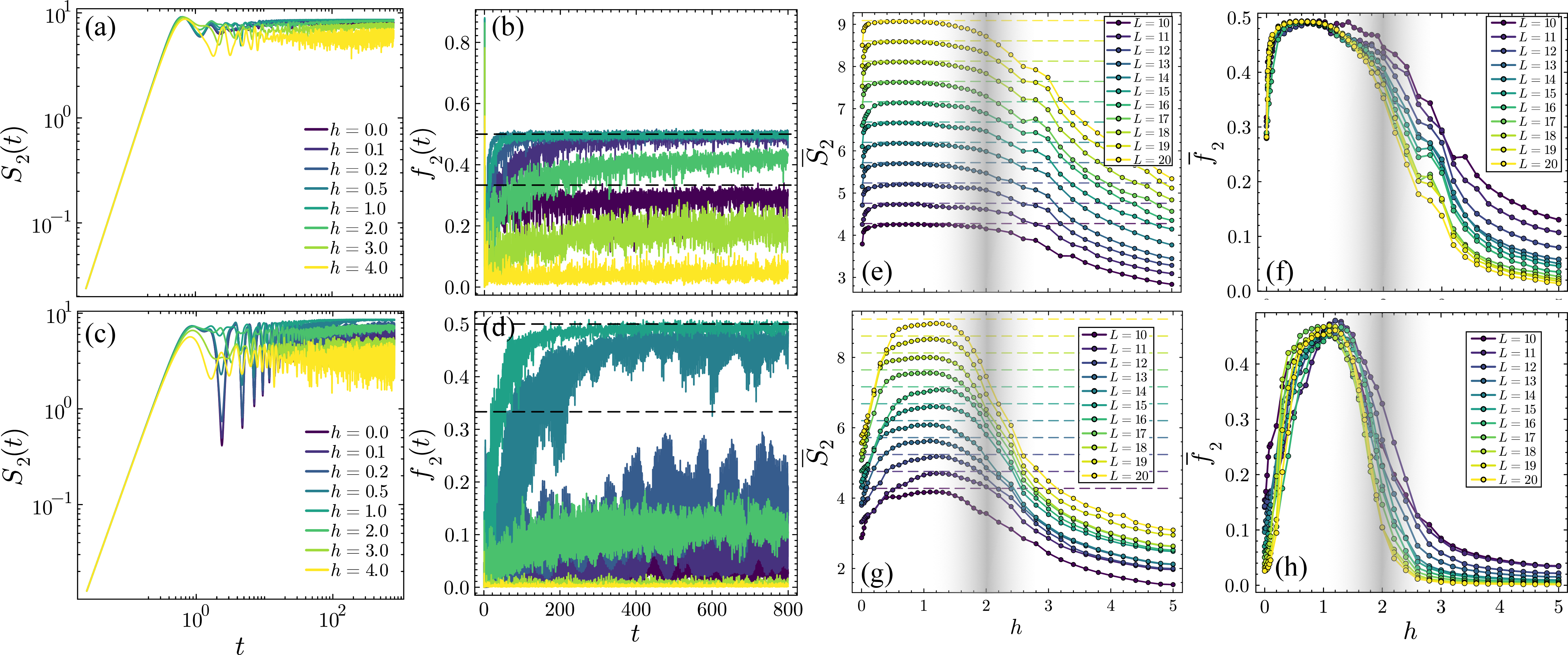}
\caption{Fock-space spreading in the presence of the quasiperiodic field
$h_j=h\cos(2\pi\beta j)$. Top row [(a), (b), (e), (f)]: polarized state
$\ket{\mathbf 0}$; bottom row [(c), (d), (g), (h)]: N\'eel state
$\ket{\mathbb Z_2}$.
(a),(c) Participation entropy $S_2(t)$ for $L=19$ and several values of $h$
(log-log scale).
(b),(d) Participation fraction $f_2(t)$ for the same parameters. Dashed lines
mark the random-state references $f_2^{\rm GUE}=1/2$ (complex amplitudes) and
$f_2^{\rm GOE}=1/3$ (real amplitudes).
(e),(g) Time-averaged entropy $\overline S_2$ and (f),(h) time-averaged
participation fraction $\overline f_2$, with $t_f=800$, versus $h$ for
$L=10,\dots,20$. In (e),(g), dashed lines of the same color mark the
random-state reference $S_2^{\rm GUE}=\ln d_{\mathcal B}-\ln2$ for each $L$
[Eq.~\eqref{eq:GUE_ergodic}]. The shaded region marks the crossover at
$h\approx2$.}
    \label{fig:qp}
\end{figure*}
The different slopes have a qualitative consequence, most evident in the
participation fraction shown in Fig.~\ref{fig:scaling}(b). Inserting the
linear fit of $\overline S_2$ into Eq.~(\ref{eq:f_scaling}) gives
\begin{equation}
  \overline f_2\geq e^{\overline S_2-\ln d_{\mathcal B}}
  \simeq e^{-(\ln\varphi-a)L+b-c_0},
  \label{eq:fPR_scaling}
\end{equation}
where we used $\ln d_{\mathcal B}\simeq L\ln\varphi+c_0$. This lower
bound follows from Jensen's inequality, $\overline{e^g}\geq e^{\overline g}$,
applied to $f_2=e^{S_2-\ln d_{\mathcal B}}$. It becomes an equality when
temporal fluctuations of $S_2(t)$ are negligible. The two initial states
illustrate the distinct behaviors discussed in Sec.~\ref{sec:spectral}.

For $\ket{\mathbf 0}$, $\overline f_2$ does not fully saturate but drifts
slowly with $L$, remaining close to, and slightly below, $f_2^{\rm GOE}$
over the accessible sizes. Consistently with $a\simeq\ln\varphi$, the
polarized state is quasi-anticoncentrated: it spreads over a large portion
of $\mathcal B_L$, with relative population fluctuations $\sigma_p d_{\mathcal B}$
of order one, comparable to those of a random state
[Eq.~(\ref{eq:fPR_variance})], yet without establishing a size-independent
ergodic fraction. For $\ket{\mathbb Z_2}$, by contrast, $\overline f_2$ decreases steadily
with $L$. Since $a<\ln\varphi$, the N\'eel dynamics spreads over exponentially
many configurations while occupying a vanishing fraction of the BFS on
average. Its relative population fluctuations also grow with system size.
The dashed lines in Fig.~\ref{fig:scaling}(b) show the bound in
Eq.~(\ref{eq:fPR_scaling}), evaluated using the fitted values of $a$ and
$b$. The bound closely tracks the polarized-state data, consistent with its
small temporal fluctuations, whereas for the N\'eel state it generally lies
below the data because of the larger oscillations of $S_2(t)$; the discrepancy
at the smallest sizes also reflects the limited accuracy of the fit.

\section{Impact of a quasiperiodic field}
\label{sec:qp}
In tight-binding models, a quasiperiodic potential is a paradigmatic route
to localization: in the Aubry--Andr\'e model, all eigenstates of a particle
hopping on a chain are extended for $h<2J$ and localized for $h>2J$, with
$J$ the hopping amplitude~\cite{roati2008anderson, modugno2009exponential, hetenyi2025numerical, longhi2021phase, an2021interactions}. Since the PXP model
is itself a hopping problem on the BFS graph [Eq.~\eqref{eq:hopping}], it is
natural to ask how its Fock-space dynamics responds to such a potential. We
therefore add a quasiperiodic longitudinal field,
\begin{equation}
    \hat H=\hat H_{\rm PXP}+\sum_{j=1}^{L}h_j\hat N_j,
    \qquad h_j=h\cos(2\pi\beta j),
    \label{eq:Hqp}
\end{equation}
with $\beta=(\sqrt5-1)/2$ the inverse golden ratio, and compare the
response of the polarized and N\'eel states. Two questions guide the
analysis. First, the field commutes with $\hat{\mathcal C}$ and therefore
breaks the chiral symmetry of $\hat H_{\rm PXP}$
(Appendix~\ref{app:ergodic}): the amplitudes are no longer constrained to
be real or imaginary, and the ergodic reference changes from
$f_2^{\rm GOE}\simeq1/3$ to $f_2^{\rm GUE}\simeq1/2$. We ask whether, and at which field
strength, each initial state actually reaches this new reference. Second, the field adds an energy offset to each link of the BFS graph. Two
neighboring configurations $\mathbf n$ and $\mathbf m$ differ only in the
state of a single atom $j$, so that their energies differ by
$E_{\mathbf m}-E_{\mathbf n}=\pm h_j$. When
these offsets are large compared with $\Omega$, the hopping between
configurations is suppressed, and the field is expected to hinder
delocalization. We ask whether this leads to localization in Fock space,
as in tight-binding models. Figure~\ref{fig:qp} addresses both questions: the top
row follows the dynamics from $\ket{\mathbf 0}$ and the bottom row that
from $\ket{\mathbb Z_2}$, showing $S_2(t)$ and $f_2(t)$ for $L=19$ together
with their time averages as functions of $h$ for $L=10,\dots,20$.

The field does not act immediately. Since it is diagonal, it leaves
$(\Delta H)^2=    \sum_{\mathbf n\neq\mathbf n_0}
    |\langle\mathbf n|\hat H|\mathbf n_0\rangle|^2$ unchanged, and the quadratic growth of
Eq.~\eqref{eq:short_time} is the same for all $h$
[Fig.~\ref{fig:qp}(a),(c)]: the initial spreading is fixed by the BFS
connectivity alone. Differences between field strengths emerge only after this regime.

For the polarized state, the field does not change the shape of $S_2(t)$
qualitatively [Fig.~\ref{fig:qp}(a)]: after the quadratic growth, $S_2(t)$
reaches a plateau for all $h$. The value of the plateau, however, depends
on $h$, as is evident from $f_2(t)$ [Fig.~\ref{fig:qp}(b)]. At $h=0$,
$f_2(t)$ fluctuates just below $f_2^{\rm GOE}\simeq1/3$. For $h=0.1$ and
$0.2$ it slowly rises towards $f_2^{\rm GUE}\simeq1/2$, and for $h=0.5$ and
$1$ it reaches this value quickly and weakly fluctuates around it. For larger
fields, $f_2(t)$ decreases again, down to $\simeq0.4$ at $h=2$, $\simeq0.2$
at $h=3$, and a few percent at $h=4$.

For the N\'eel state, the field changes the dynamics qualitatively
[Fig.~\ref{fig:qp}(c),(d)]. At $h=0.1$ and $0.2$, $S_2(t)$ still dips at
the revival times, and $f_2(t)$ remains small, with pronounced beatings: the
scarred dynamics survives the weak field, and $f_2(t)$ stays far from both
random-state references. At $h=0.5$, the dips are progressively washed
out, and $f_2(t)$ approaches $1/2$ over $t\sim200$, with sporadic drops.
At $h=1$, $f_2(t)$ reaches $1/2$ within a few tens of time units and
fluctuates around it. For larger fields, $f_2(t)$ drops again, and faster
than for the polarized state: it fluctuates around $0.1$ at $h=2$ and
remains below a few percent for $h\geq3$.

To determine how these behaviors scale with system size, we consider the
time averages $\overline S_2$ and $\overline f_2$ as functions of $h$.
For $\ket{\mathbf 0}$ [Fig.~\ref{fig:qp}(e),(f)], $\overline f_2$ rises sharply from $\simeq0.3$ at $h=0$ to $\simeq 1/2$ at $h\approx0.3$, and
remains on this plateau up to $h\approx1.3$. On the plateau, the participation fractions for all $L$ collapse,
signaling anticoncentration: $\overline f_2\simeq1/2$ independently of the
system size, and $\overline S_2$ lies on the reference
$S_2^{\rm GUE}\simeq\ln d_{\mathcal B}-\ln2$ (dashed lines), so that half
of the BFS is effectively occupied at every size. Since the dynamics was
already spreading over the whole BFS at $h=0$, the increase from $1/3$ to
$1/2$ is entirely due to the change in the statistics of the amplitudes.
Increasing $h$ further, the system
undergoes a crossover, centered at $h\approx2$ (shaded region), between two
distinct regimes of Fock-space delocalization. Below the crossover,
$\overline f_2$ is independent of $L$. Across it, the curves for different
$L$ separate, and above it $\overline f_2$ decreases systematically with
$L$, while $\overline S_2$ falls below the GUE reference by an amount that
grows with $L$. $\overline S_2$ nevertheless keeps growing linearly with
$L$, with a reduced slope, $a=0.26$ at $h=5$ implying $\overline{\mathcal{D}}_2 \approx 0.54$. The crossover thus separates a regime in which
the dynamics occupies a finite fraction of the BFS from one in which it
occupies exponentially many configurations, but a vanishing fraction of
the BFS.

For $\ket{\mathbb Z_2}$ [Fig.~\ref{fig:qp}(g),(h)], the behavior is
different at weak field. At $h=0$, $\overline f_2$ is small and decreases
with $L$, as discussed in Sec.~\ref{sec:spectral_dynamics}, and it grows
only gradually with $h$. It reaches its maximum, $\overline f_2\lesssim1/2$, at $h\approx1$--$1.5$,
where it depends only weakly on $L$: the dynamics is close to
anticoncentration, and $\mathcal D_2\lesssim 1$. The window of maximal delocalization is narrower than for
the polarized state, and beyond the crossover region $\overline f_2$
decreases with $L$ as for $\ket{\mathbf 0}$. $\overline S_2$ again grows in $L$, without showing localization.

These results are summarized in
Table~\ref{tab:summary}. First, both initial states reach the
delocalization of complex random states,
$\overline f_2\simeq f_2^{\rm GUE}$, but at different field strengths. For
the polarized state an arbitrarily weak field is sufficient, since breaking
the chiral symmetry only changes the statistics of the amplitudes. For the
N\'eel state a field of order $\Omega$ is needed. In the window
$h\approx1$, scarred and thermalizing initial states thus become
indistinguishable in terms of Fock-space delocalization. Second, a strong
field does not localize the dynamics within the accessible sizes:
$\mathrm{PR}_2$ keeps growing exponentially with $L$, and, unlike in the
non-interacting Aubry--Andr\'e model, we find no sign of a sharp transition. Strong fields
instead produce a partial delocalization, with $0<\mathcal D_2<1$, similar
to that of the scarred dynamics at $h=0$ but with a different origin: the
suppression of hopping by the energy offsets, rather than non-thermal
eigenstates. Read along its rows, Table~\ref{tab:summary} shows that the
field drives both initial states through the same sequence, ending in a
vanishing fraction of the BFS, but starting from different points: the
polarized state is already close to anticoncentration at $h=0$, while the
N\'eel state is not. Read along its columns, it shows that the two states
differ only at $h=0$, where the scars set them apart, and behave alike
both in the window $h\approx1$ and beyond the crossover.
\begin{table}[!t]
\caption{Fock-space delocalization of the dynamics from the polarized and
N\'eel states, for different strengths $h$ of the quasiperiodic field.}
\label{tab:summary}
\begin{ruledtabular}
\begin{tabular}{|l|c|c|c|}
 & $h=0$ & $h\approx1$ & $h>2$ \\
\hline
$\ket{\mathbf 0}$
 & \begin{tabular}[c]{@{}c@{}}quasi-\\anticoncentrated\\ ($f_2\simeq1/3$, GOE)\end{tabular}
 & \begin{tabular}[c]{@{}c@{}}anticoncentrated\\ ($f_2\simeq1/2$, GUE)\end{tabular}
 & \begin{tabular}[c]{@{}c@{}}delocalized,\\ vanishing fraction\end{tabular} \\[12pt]
$\ket{\mathbb Z_2}$
 & \begin{tabular}[c]{@{}c@{}}delocalized,\\ vanishing fraction\end{tabular}
 & \begin{tabular}[c]{@{}c@{}}quasi-\\anticoncentrated\\ ($f_2\simeq1/2$, GUE)\end{tabular}
 & \begin{tabular}[c]{@{}c@{}}delocalized,\\ vanishing fraction\end{tabular} \\
\end{tabular}
\end{ruledtabular}
\end{table}

\section{Conclusions and outlook}
We studied how the wave function of the PXP model spreads over the
blockaded Fock space, using the participation entropy $S_2$ and the
participation fraction $f_2$, and how this spreading is modified by a
quasiperiodic longitudinal field.

Without field, neither the scars nor the scarred dynamics are localized.
The scars occupy a number of configurations that grows exponentially with
$L$, but more slowly than the dimension of the BFS, so that the occupied
fraction vanishes at large sizes ($\mathcal D_2\simeq0.64$). The other
eigenstates are instead close to the random-state reference with real
amplitudes, $f_2^{\rm GOE}\simeq1/3$. The dynamics reflects this picture. Starting from the polarized
state, the wave function relaxes to a fraction close to $1/3$ that depends
only weakly on $L$ ($\mathcal D_2\simeq0.98$), close to anticoncentration.
Starting from the N\'eel state, it periodically occupies a sizable
fraction of the BFS between revivals, but its time-averaged fraction
decays with $L$ ($\mathcal D_2\simeq0.54$): the scarred dynamics is
delocalized without being ergodic.

The quasiperiodic field plays two roles. First, it breaks the chiral
symmetry, so that the amplitudes are no longer constrained to be real or
imaginary and the random-state reference rises from $1/3$ to $1/2$.
Second, it detunes the links of the BFS graph and thus controls the
delocalization. For the polarized state, an arbitrarily weak field brings
the occupied fraction to $1/2$, independently of $L$: the dynamics is robustly
anticoncentrated. For the N\'eel state, the field must be of order the PXP coupling to destroy scarring dynamics. Then the fraction also approaches $1/2$, almost independently of
$L$, but only in a narrower range of $h$. In this window, N\'eel and polarized initial
states become indistinguishable in terms of occupied fraction. Beyond a crossover at
$h\approx2$, both initial states lose anticoncentration: $S_2$ still grows
linearly with $L$, but the occupied fraction decreases with $L$, and no
localization is observed at the accessible sizes.

These results show that Fock-space delocalization provides a sharper
characterization of weak ergodicity breaking than the return probability
alone, and that a simple quasiperiodic field can tune the PXP dynamics
between different delocalization regimes. These quantities are accessible in Rydberg-atom arrays. Their estimation, however,
requires a number of snapshots that grows exponentially with $L$, which
restricts this approach to small system sizes. Since anticoncentration is
a key requirement of sampling-based protocols, we hope that the control
offered by the quasiperiodic field may be useful for designing such
protocols with analog quantum simulators. Natural extensions include 
the exploration of other initial states, fields, and scarred
models, the full multifractal spectrum $\mathcal D_q$, and, as a more
ambitious goal, tracking not only how much of the BFS the wave function
occupies, but where on the graph its weight goes. On the applied side,
the control of Fock-space spreading offered by the quasiperiodic field
could be exploited in quantum machine learning, where the measured
configuration probabilities serve as an expanded feature
space~\cite{delorenzis2025harnessing, delorenzis2026entanglement, mujal2021opportunities, perciavalle2025predicting, innocenti2023potential}, and the participation fraction sets how many of
these features are effectively populated.

\section{Acknowledgements}
We thank A. Palamara, N. Lo Gullo and A. De Lorenzis for fruitful discussions. The author used Claude (Anthropic) and ChatGPT (OpenAI) for assistance with coding and editing. All outputs were checked by the author, who takes full responsibility for the content. We acknowledge support from PNRR MUR Project No. PE0000023-NQSTI through the secondary project ThAnQ.

\appendix

\section{Short-time behavior of participation entropies}
\label{app:short-time}
Consider an initial Fock state $|\mathbf n_0\rangle$ evolving under
a time-independent Hamiltonian \[\hat H = \sum_{\mathbf n, \mathbf m}\langle\mathbf n|\hat H|\mathbf m\rangle\ket{\mathbf{n}}\bra{\mathbf{m}}.\] At short
times, the probability of remaining in the initial configuration is
\begin{equation}
    |C_{\mathbf n_0}(t)|^2
    \simeq 1-At^2,
\end{equation}
where 
\begin{equation}
    A
    =
    \sum_{\mathbf n\neq\mathbf n_0}
    |\langle\mathbf n|\hat H|\mathbf n_0\rangle|^2
    =
    \langle\hat H^2\rangle_0-\langle\hat H\rangle_0^2
    =
    (\Delta H)^2.
\end{equation}
For the remaining configurations ($\mathbf{n}\neq \mathbf{n}_0$),
\begin{equation}
    |C_{\mathbf n}(t)|^2
    \simeq
    t^2|\langle\mathbf n|\hat H|\mathbf n_0\rangle|^2.
\end{equation}
Consequently, for fixed $q>1$, the generalized inverse participation
ratio satisfies
\begin{equation}
    I_q(t)
    =
    \sum_{\mathbf n\in\mathcal B_L}|C_{\mathbf n}(t)|^{2q}
    \simeq 1-q(\Delta H)^2t^2.
\end{equation}
Using $S_q(t)=(1-q)^{-1}\ln I_q(t)$ and expanding the logarithm,
we obtain
\begin{equation}
    S_q(t)\simeq
    \frac{q}{q-1}(\Delta H)^2t^2.
\end{equation}
The growth of the participation entropy is governed by the energy
uncertainty $\Delta H$, which also determines the Mandelstam--Tamm
(MT) quantum speed limit, a lower bound on the time required for a
quantum state to evolve into an orthogonal state,
$\tau_{\mathrm{MT}}=\pi/(2\Delta H)$ ($\tau_{\perp}\geq \tau_{\rm MT}$ with $\tau_{\perp}$ being the orthogonalization time)~\cite{deffner2017quantum, perciavalle2024quantum, mandelstam1991uncertainty, giovannetti2003quantum}.
For $q=2$, the participation entropy exhibits quadratic short-time
growth,
\begin{equation}
    S_2(t)\simeq 2(\Delta H)^2t^2,
\end{equation}
as long as the evolved state remains close to its initial
configuration. In particular, the logarithmic expansion requires
$2(\Delta H)^2t^2\ll1$, or equivalently,
\begin{equation}
    t\ll\frac{1}{\sqrt{2}\Delta H}
    =\frac{\sqrt{2}}{\pi}\tau_{\mathrm{MT}}.
\end{equation}
This condition identifies the regime of validity of the quadratic
approximation and relates it to the MT timescale. Note that for general $q$ the condition reads $t\ll \frac{2}{\sqrt{q}}\frac{\tau_{\rm MT}}{\pi}$.

\section{Ergodic reference value of the participation ratio}
\label{app:ergodic}

\paragraph*{Reality of the amplitudes.}
The PXP model has chiral symmetry, i.e., the operator $\hat{\mathcal C}=\prod_{j}\hat Z_j$ ($\hat{\mathcal C}^2=\hat 1$)
anticommutes with $\hat H_{\mathrm{PXP}}$: every term, including the boundary
ones, contains exactly one $\hat X_j$, and $\hat Z\hat X\hat Z=-\hat X$, while
$\hat Z\hat P\hat Z=\hat P$. Hence
$\hat{\mathcal C}\hat H^k_{\mathrm{PXP}}\hat{\mathcal C}
=(-1)^k\hat H^k_{\mathrm{PXP}}$ and
\begin{equation}
  \hat{\mathcal C}\,e^{-i\hat H_{\mathrm{PXP}}t}\,\hat{\mathcal C}
  =e^{i\hat H_{\mathrm{PXP}}t}.
  \label{eq:CUC}
\end{equation}
Fock states satisfy $\hat{\mathcal C}|\mathbf n\rangle
=s_{\mathbf n}|\mathbf n\rangle$, with $s_{\mathbf n}=(-1)^{L-N_{\mathbf n}}$
and $N_{\mathbf n}=\sum_j n_j$ being the total number of excitations of the Fock state. Moreover, the matrix elements
$\langle\mathbf n|\hat H_{\mathrm{PXP}}|\mathbf m\rangle$ are either $\Omega$
or $0$; inserting resolutions of the identity, every
$\langle\mathbf n|\hat H^k_{\mathrm{PXP}}|\mathbf n_0\rangle$ is a sum of
products of real numbers, hence real. Conjugating the series of
$C_{\mathbf n}(t)$ term by term and using Eq.~\eqref{eq:CUC},
\begin{equation}
\begin{split}
  C_{\mathbf n}(t)^*
  &=\langle\mathbf n|e^{i\hat H_{\mathrm{PXP}}t}|\mathbf n_0\rangle
   =\langle\mathbf n|\hat{\mathcal C}e^{-i\hat H_{\mathrm{PXP}}t}
    \hat{\mathcal C}|\mathbf n_0\rangle\\
  &=(-1)^{N_{\mathbf n}-N_{\mathbf n_0}}\,C_{\mathbf n}(t).
\end{split}
\end{equation}
Thus $C_{\mathbf n}$ is real (purely imaginary) when
$N_{\mathbf n}-N_{\mathbf n_0}$ is even (odd):
\begin{equation}
  C_{\mathbf n}(t)=i^{\,\delta_{\mathbf n}}\,r_{\mathbf n}(t),\qquad
  r_{\mathbf n}(t)\in\mathbb R,
  \label{eq:real}
\end{equation}
with $\delta_{\mathbf n}=(N_{\mathbf n}-N_{\mathbf n_0})\bmod 2$, so that
$p_{\mathbf n}=|C_{\mathbf{n}}|^2=r_{\mathbf n}^2$ is controlled by a single real variable.

\paragraph*{Ergodic reference.}
Typical eigenstates of chaotic systems, as well as states evolved under
chaotic dynamics, are well described in the thermodynamic limit by random
states, whose amplitudes in a generic basis behave as independent Gaussian
variables with zero mean and variance $1/d_{\mathcal B}$~\cite{Lakshminarayan2026}.
For random states with real amplitudes one has results consistent with Gaussian Orthogonal Ensemble (GOE)~\cite{Lakshminarayan2026}
\begin{equation}
  \mathbb{E}[I_2]=I_2^{\rm GOE}\simeq \frac{3}{d_\mathcal{B}},
  \qquad
  f_2^{\rm GOE}\simeq \frac{1}{3},
\end{equation}
whereas for complex amplitudes results are consistent with the Gausssian Unitary Ensemble (GUE)~\cite{Lakshminarayan2026}
\begin{equation}
  \mathbb{E}[I_2]=I_2^{\rm GUE}\simeq \frac{2}{d_\mathcal{B}},
  \qquad
  f_2^{\rm GUE}\simeq \frac{1}{2}.
\end{equation}
Here
$\mathbb{E}[\ldots]$ denotes the statistical average,
and the approximate equalities become exact in the thermodynamic limit
$d_{\mathcal{B}}\rightarrow\infty$.

In the PXP model, Eq.~\eqref{eq:real} holds, so the phase factor
$i^{\,\delta_{\mathbf n}}$ drops out of the modulus and
$|C_{\mathbf n}|^4 = r_{\mathbf n}^4$. Taking the real amplitudes
$r_{\mathbf n}$ to be Gaussian with variance $1/d_{\mathcal B}$, we obtain
\begin{equation}
  \mathbb{E}[|C_{\mathbf n}|^4]
  = \mathbb{E}[r_{\mathbf n}^4] \simeq \frac{3}{d_{\mathcal B}^2},
\end{equation}
and therefore
\begin{equation}
  \mathbb{E}[I_2]
  = \sum_{\mathbf n}\mathbb{E}[|C_{\mathbf n}|^4]
  \simeq \frac{3}{d_{\mathcal B}}.
\end{equation}
The ergodic reference for the PXP dynamics is thus
\begin{equation}
  f_2^{\rm GOE}\simeq \frac{1}{3},
  \qquad
  S_2^{\rm GOE}\simeq\ln d_{\mathcal B}-\ln 3 .
\end{equation}

For comparison, generic complex amplitudes
$C_{\mathbf n}=x_{\mathbf n}+i\,y_{\mathbf n}$, with $x_{\mathbf n}$ and
$y_{\mathbf n}$ independent real Gaussian variables of zero mean and
variance $1/(2d_{\mathcal B})$ each, yield instead
\begin{equation}
  \mathbb{E}[|C_{\mathbf n}|^4] \simeq \frac{2}{d_{\mathcal B}^2},
\end{equation}
hence
\begin{equation}
  \mathbb{E}[I_2]\simeq \frac{2}{d_{\mathcal B}},
  \qquad
  f_2^{\rm GUE}\simeq \frac{1}{2},
  \qquad
  S_2^{\rm GUE}\simeq\ln d_{\mathcal B}-\ln 2 .
\end{equation}
The populations $p_{\mathbf n}=|C_{\mathbf n}|^2$ are consistent with the
Porter--Thomas distribution~\cite{claeys2025fock}.

\paragraph*{Breaking the chiral symmetry.}
A longitudinal field $\hat V=\sum_j h_j\hat N_j$ commutes with
$\hat{\mathcal C}$, so that
$\hat{\mathcal C}(\hat H_{\mathrm{PXP}}+\hat V)\hat{\mathcal C}
=-\hat H_{\mathrm{PXP}}+\hat V$. Eq.~\eqref{eq:real} then no longer holds,
and, for ergodic dynamics, the reference is expected to cross over from GOE to GUE.

\section{Ergodic reference values of the participation entropies}
\label{app:ergodic_q}

We extend the ergodic reference of Appendix~\ref{app:ergodic} to integer
$q\geq2$. By Eq.~\eqref{eq:real}, for an ergodic state the real amplitudes
$r_{\mathbf n}$ behave as independent Gaussian variables with zero mean and
variance $1/d_{\mathcal B}$. Their moments are
$\mathbb{E}[r_{\mathbf n}^{2q}]\simeq(2q-1)!!\,d_{\mathcal B}^{-q}$, so that,
to leading order in $1/d_{\mathcal B}$,
\begin{equation}
  \mathbb{E}[I_q]=\sum_{\mathbf n}\mathbb{E}[r_{\mathbf n}^{2q}]
  \simeq (2q-1)!!\,d_{\mathcal B}^{\,1-q},
  \label{eq:Iq_erg}
\end{equation}
which reduces to $\mathbb{E}[I_2]\simeq3/d_{\mathcal B}$ for $q=2$.
Since the relative fluctuations of $I_q$ vanish for
$d_{\mathcal B}\to\infty$, the participation entropies are
\begin{equation}
  S_q^{\mathrm{GOE}}=\ln d_{\mathcal B}-\frac{\ln(2q-1)!!}{q-1},
  \label{eq:Sq_erg}
\end{equation}
e.g., $S_2^{\mathrm{GOE}}=\ln d_{\mathcal B}-\ln3$,
$S_3^{\mathrm{GOE}}=\ln d_{\mathcal B}-\tfrac12\ln15$, and
$S_4^{\mathrm{GOE}}=\ln d_{\mathcal B}-\tfrac13\ln105$.

Using $d_{\mathcal B}=F_{L+2}\approx\varphi^{L+2}/\sqrt5$, we obtain
\begin{equation}
  S_q^{\mathrm{GOE}}(L)\approx L\ln\varphi+\left(2\ln\varphi-\tfrac12\ln5\right)
  -\frac{\ln(2q-1)!!}{q-1},
  \label{eq:Sq_erg_L}
\end{equation}
with $\ln\varphi\simeq0.481$ and $2\ln\varphi-\tfrac12\ln5\simeq0.158$.

\end{document}